\documentclass[fleqn,usenatbib]{mnras}

\usepackage{newtxtext,newtxmath}
\usepackage{threeparttable}
\usepackage{academicons}

\DeclareRobustCommand{\VAN}[3]{#2}

\let\VANthebibliography\thebibliography
\def\thebibliography{\DeclareRobustCommand{\VAN}[3]{##3}\VANthebibliography}

\usepackage[T1]{fontenc}
\usepackage{hyperref}
\usepackage{wrapfig}
\usepackage{amsfonts}
\usepackage{amsmath}
\usepackage{tablefootnote}
\usepackage{bm}
\usepackage{graphicx}
\usepackage{epsfig}
\usepackage{mathtools}
\usepackage{color}
\usepackage{float}
\usepackage{lipsum}
\usepackage{booktabs}
\usepackage{orcidlink}
\usepackage{placeins}

\usepackage{graphicx}	
\usepackage{amsmath}	
\usepackage[capitalize]{cleveref}
\usepackage{breqn}
\usepackage[mathscr]{euscript}
\usepackage{gensymb}
\usepackage{cuted}
\usepackage{widetext}
\usepackage{enumitem}
\usepackage{minibox}

\newcommand{\pplus}{Pantheon$+$}
\newcommand{\LA}{\Lambda}
\newcommand{\LCDM}{$\LA$CDM}
\newcommand{\lcdm}{spatially flat $\LA$CDM}

\newcommand{\sne}{SNe~Ia}
\newcommand{\laneetal}{\citet{Lane_2023}}
\newcommand{\diff}{\mathrm{d}}
\newcommand{\prior}{\mathrm{Pr}}
\newcommand{\goesas}{\mathop{\sim}\limits}

\newcommand{\Z}[1]{_{\lower2pt\hbox{$\scriptstyle#1$}}}
\newcommand{\X}[1]{_{\lower1pt\hbox{$\scriptscriptstyle#1$}}}
\newcommand{\Ns}[1]{_{\lower2pt\hbox{$\scriptstyle\rm#1$}}}

\newcommand{\w}[1]{\mathop{\hbox{\,#1}}}
\newcommand{\lsim}{\lesssim}
\newcommand{\gsim}{\gtrsim}

\newcommand{\h}{\,h^{-1}}
\newcommand{\hM}{\h\w{Mpc}}

\newcommand{\ns}[1]{_{\rm #1}}

\newcommand{\Hn}{H\Z0}
\newcommand{\dL}{d\Z{L}}

\newcommand{\zmin}{z\ns{min}}

\newcommand{\beq}{\begin{equation}}
\newcommand{\eeq}{\end{equation}}
\newcommand{\hblank}[1]{\hbox to#1 mm{\hfil}}

{
    \definecolor{BLACK}{gray}{0}
    \definecolor{WHITE}{gray}{1}
    \definecolor{RED}{rgb}{1,0,0}
    \definecolor{GREEN}{rgb}{0,1,0}
    \definecolor{dgreen}{rgb}{.1,.6,.1}
    \definecolor{BLUE}{rgb}{0,0,1}
    \definecolor{CYAN}{cmyk}{1,0,0,0}
    \definecolor{MAGENTA}{cmyk}{0,1,0,0}
    \definecolor{YELLOW}{cmyk}{0,0,1,0}
    \definecolor{aw}{rgb}{0.2,0.5,0.75}
}
\hypersetup{
    colorlinks,%
    citecolor=blue,%
    filecolor=blue,%
    linkcolor=blue,
    urlcolor=blue
}

\definecolor{MyDarkRed}{rgb}{0.71,0.14,0.07}

\definecolor{ABlue}{rgb}{0.0,0.7,0.8}

\definecolor{MyGreen}{rgb}{0.0,0.5,0.0}

\definecolor{amethyst}{rgb}{0.6, 0.4, 0.8}

\definecolor{MyOrange}{rgb}{0.93, 0.53, 0.18}

\title[Evidence for fundamental change in cosmology]{Supernovae evidence for foundational change to cosmological models}

\author[Seifert et al.]{\fontsize{13pt}{16pt}\selectfont Antonia Seifert$^{1,2}$\orcidlink{0009-0005-9892-3667}, Zachary G.~Lane$^{1}$\orcidlink{0009-0003-8380-4003}\thanks{Corresponding author\newline zachary.lane@pg.canterbury.ac.nz}, Marco Galoppo$^1$\orcidlink{ 0000-0003-2783-3603}\newauthor \fontsize{13pt}{16pt}\selectfont
Ryan Ridden-Harper$^{1}$\orcidlink{0000-0003-1724-2885},
and David L.~Wiltshire$^{1}$\orcidlink{0000-0003-1992-6682}
\\
$^{1}$School of Physical and Chemical Sciences — Te Kura Mat\={u}, University of Canterbury, Private Bag 4800, Christchurch 8140, New Zealand\\
$^{2}$Institut f\"ur Theoretische Physik, Universit\"at Heidelberg, Philosophenweg 12, D-69120 Heidelberg, Germany\\ 
}

\date{Accepted 2024 November 21. Received 2024 November 20; in original form 2024 July 11}

\pubyear{2024}

\begin{document}
\label{firstpage}
\pagerange{\pageref{firstpage}--\pageref{lastpage}}
\maketitle

\begin{abstract}

We present a new, cosmologically model-independent, statistical analysis of the \pplus\ type Ia supernovae spectroscopic dataset, improving a standard methodology adopted by Lane et al. We use the Tripp equation for supernova standardisation alone, thereby avoiding any potential correlation in the stretch and colour distributions. We compare the standard homogeneous cosmological model, i.e., \lcdm, and the timescape cosmology which invokes backreaction of inhomogeneities. Timescape, while statistically homogeneous and isotropic, departs from average Friedmann-Lema\^{\i}tre-Robertson-Walker evolution, and replaces dark energy by kinetic gravitational energy and its gradients, in explaining independent cosmological observations. When considering the entire \pplus\ sample, we find very strong evidence ($\ln B> 5$) in favour of timescape over \LCDM. Furthermore, even restricting the sample to redshifts beyond any conventional scale of statistical homogeneity, $z > 0.075$, timescape  is preferred over \LCDM\ with $\ln B> 1$. These results provide evidence for a need to revisit the foundations of theoretical and observational cosmology.
\end{abstract}



\section{Introduction}\label{sec:intro}

The $\Lambda$ Cold Dark Matter (\LCDM) model, which has served as 
the standard cosmological model for quarter of a century, is facing serious challenges in light of recent results \citep{DES_2024, DESI_2024} and may need to be reconsidered at a fundamental level \citep{Di_Valentino_2021, Peebles_2022, Aluri_2022}. In this Letter, we present definite statistical evidence that the timescape cosmological model \citep{Wiltshire_2007_clocks,Wiltshire_2007_sol,Wiltshire_2009_obs} outperforms \LCDM\ in matching Type Ia Supernovae (\sne) observations. It may provide not only a viable alternative to the standard cosmological model, but ultimately a preferred one. This result potentially has far-reaching consequences not only for cosmology, but also for other key aspects of astrophysical modelling from late epochs to the early universe. 

We perform an empirical cosmologically independent analysis within which both the \LCDM\ and timescape cosmologies may be embedded, and thus compared via Bayesian statistics. The timescape model is a particular implementation of Buchert’s scalar averaging scheme which incorporates backreaction of inhomogeneities \citep{Buchert_2000, Buchert_2001, Buchert_2020, Wiltshire_2014_cosmic}. Instead of a matter density parameter relative to average Friedmann-Lema\^{\i}tre-Robertson-Walker model (as in $\Lambda$CDM), timescape is characterised by the \textit{void fraction}, $f\Ns{v}$, which represents the fractional volume of the expanding regions of the universe made up by voids.

A key ingredient of the timescape model is a particular integrability relation for the Buchert equations: the {\em uniform quasilocal Hubble expansion} condition. Physically, it is motivated by an extension of Einstein's Strong Equivalence Principle to cosmological averages at small scales ($\goesas4\,$--$\,15\,$Mpc) where perturbations to average isotropic expansion and average isotropic motion cannot be observationally distinguished \citep{Wiltshire_2008}. 

In standard cosmology, differences from average FLRW expansion are assumed to be mostly attributed to local Lorentz boosts --- i.e., peculiar velocities --- of source and observer, with gravitational potentials contributing fractional variations of $\goesas10^{-5}$ of average expansion at galaxy and galaxy cluster scales. In timescape, the same fractional variation can be up to $\goesas10^{-3}$ and the equivalence of different choices of background, via the Cosmological Equivalence Principle, means that notions of average isotropic expansion persist well into the nonlinear regime of structure formation. The signature of the emergent kinetic spatial curvature of voids has now been identified in cosmological simulations using full numerical general relativity without $\Lambda$ \citep{Williams_2024}. 

Both the standard cosmology and the timescape model agree empirically on a Statistical Homogeneity Scale (SHS), typically given as $z \Ns {CMB} \goesas 0.033$ by the two--point galaxy correlation function \citep{Hogg_2005, Scrimgeour_2012, Dam_2017}. Timescape offers its most important tests and predictions below the SHS, at scales where the filaments, sheets and voids of the cosmic web are still expanding but in the nonlinear regime.

To conduct our analysis, we use the largest spectroscopically confirmed \sne\ dataset, \pplus\ \citep{Scolnic_2022}. \sne\ have been a pillar for informing the distance ladder used for cosmological model comparison, and have a rich history in revolutionising the field \citep{Riess_1998, Perlmutter_1999}. More modern methods for standardising \sne\ light-curves use the SALT2 fitting algorithm \citep{Guy_2007, Taylor_2021}, as used by \pplus, and more recently SALT3 \citep{Kenworthy_2021} used by the Dark Energy Survey 5-year release \citep[DES5yr,][]{DES_2024}. The SALT fitting algorithms fit the distance moduli, $\mu$, using a modified version
of the Tripp formula:
\begin{equation}\label{eq:tripp}
    \mu = m\Ns{B}^* - M\Ns{B} + \alpha x\Z 1 - \beta c\,,
\end{equation}
where $\alpha$ and $\beta$ are considered constant across all redshifts\footnote{For \pplus, \citet{Scolnic_2022} adopt values of $\alpha = 0.148$ and $\beta = 3.112$, respectively, for their nominal fit.}, $x\Z 1$ is the time stretch/decay parameter, $c$ is the colour, and $m\Ns{B}$ and $M\Ns{B}$ are the apparent and absolute magnitude in the rest-frame of the $B$~band filter. Rest-frame measurements are identical for theories obeying the Strong Equivalence Principle of general relativity -- in particular, in both the FLRW and timescape models. In our analysis, $x\Z 1$, $c$, and $m\Ns{B}$ are taken directly from the \pplus\ data.

The observational distance modulus from \cref{eq:tripp} is then compared with the theoretical distance modulus, given by
\begin{equation}
    \mu \equiv 25 + 5 \log_{10}\left(\frac{\dL}{\rm Mpc}\right)\label{eq:mu}\,,
\end{equation}
which is determined using the bolometric flux. The luminosity distance, $\dL$, can be calculated using the redshift of the supernovae and suitable cosmological model parameters. Typically, these are $\Omega\Ns{M0}$ for the \lcdm\ model and $f\Ns{v0}$ for the timescape cosmology.\footnote{See \citet[Appendix~A]{Dam_2017} for detailed comparisons of luminosity distance calculations in the timescape and FLRW models.} Thus, the distance modulus constitutes the pillar of cosmological model comparison via supernovae analysis.

As noted in \citet{Lane_2023}, we omit peculiar velocity corrections. These are typically made using FLRW geometry assumptions, making it impossible to include them while preserving model-independence, or to perform a fair comparison. However, as distinctions between peculiar motion and expansion are central to the further development of timescape, the inclusion of such corrections will be addressed in future work. We would expect such corrections to have a small impact for low-redshift data cuts and negligible impact for $\zmin$ taken within a statistically homogeneous regime \citep{Carr_2022}. Furthermore, for the same reasons we do not include other cosmological model and metric-dependent bias corrections, such as Malmquist biases. Such corrections are small and cannot drive any substantial changes to the Bayes factors since the trend with redshift is expected to be very similar\footnote{The principal small difference occurs in the geometric homogeneous Eddington bias \citep{McKay_thesis_2016}, leading to the potential for future tests.} in both \LCDM\ and timescape.

\citet{Lane_2023} already presented moderate preference in favour of the timescape model over \LCDM. A similar result was also obtained by the DES team, with $\zmin=0.033$, using the Akaike Information Criterion (AIC) on the DES5yr supernovae sample \citep{Camilleri_2024}. They further noted a change from $\frac{1}{2}\Delta {\rm AIC} = -1.7$ (in favour of timescape) to $\frac{1}{2}\Delta {\rm AIC} = 6.3$ (in favour of \lcdm), when \sne \, data were combined with Baryonic Acoustic Oscillation (BAO) measurements. 
However, the BAO analysis of \citet{Camilleri_2024} assumes purely geometric adjustments to the standard FLRW pipeline, using a \LCDM\ calibration of the BAO drag epoch, which is not the case in timescape. Incorporating detailed BAO analysis into the timescape cosmology requires extraction of the BAO from galaxy clustering statistics, which has already been implemented \citep{Heinesen_2019}. However, since the ratio of baryonic matter to nonbaryonic dark matter is different from \LCDM, matter model calibrations in the early universe must also be revisited.

\section{Statistical Analysis}\label{sec:statmethods}

We determine Bayes factors, $B$, using the standard Jeffrey's scale \citep{Kass_1995} for model comparison, whereby $| \ln{B} | < 1$ indicates no statistical preference, $1 \leq | \ln{B} | < 3$ moderate preference, while $3 \leq | \ln{B} | < 5$ and $| \ln{B} | \geq 5$ represent strong and very strong preference respectively. In this Letter, positive (negative) $\ln B$ values indicate a preference for the timescape (\lcdm) model.

Bayesian statistics have already been implemented on \sne\ data for cosmological analysis, originally in the SDSS one-year sample \citep{Kessler_2009_Sloan, March_2011} but later extended to the Joint Lightcurve Analysis \citep[JLA,][]{Betoule_2014} sample \citep{Nielsen_2016, Dam_2017} and more recently in the \pplus\ \citep{Scolnic_2022, Brout_2022_cosmo, Brout_2022_cal} dataset \citep{Lane_2023}. 

The previous studies implemented a Bayesian hierarchical likelihood construction in the form
\begin{align}\label{eq:likelihood}
    \mathcal{L} &\equiv \prod_{i = 1}^N \prior\left[\left.(\hat{m}\Ns{B}^{*}, \hat{x}\Z 1, \hat{c}) \Z i \right| H\right] \nonumber\\
    &=  \prod_{i = 1}^N \int \prior\left[\left.(\hat{m}\Ns{B}^{*}, \hat{x}\Z 1, \hat{c}) \Z i \right| (M\Ns{B},x\Z 1,c) \Z i, H\right] \nonumber\\
    &\hblank{8} \times \prior\left[\left.(M\Ns{B}, x\Z 1, c) \Z i \right| H\right]  \diff M\Ns{B} \diff x\Z 1 \diff c,
\end{align}
where the quantities which are denoted with a hat are considered to be observed values, the true values are the quantities not denoted by a hat, and $N$ is the number of supernovae observations. The true data represents the intrinsic parameters utilised explicitly in the Tripp \citep{Tripp_1998} relation.

\citet{Nielsen_2016}, \citet{Dam_2017} and \citet{Lane_2023} follow the analysis of \citet{March_2011} and adopt global, independent Gaussian distributions for $M \Ns{B}$, $x\Z 1$ and $c$ to determine the probability density of the true parameters. However, both of these simplifying assumptions are ultimately flawed. Indeed, (i) the true values of $x\Z 1$ and $c$ are expected to be highly correlated as these are effective parameters obtained by coarse-graining the highly complex processes behind supernovae explosions; (ii) both the distributions of $x\Z1$ and $c$ present strong non-Gaussian features that cannot be explained away by systematics or biases in the data. Whilst the former always represented an overly-simplifying assumption, the latter was a reasonable assumption when it was first implemented, however, the vast increases in observed \sne\ have shown the second assumption to be flawed \citep{Hinton_2019}. 

To overcome the faulty assumptions of the previous analyses, a full non-Gaussian modelling of the joint distribution for $x\Z1$ and $c$ would be required. This represents non-trivial changes in the likelihood construction and integration, which will be addressed in future work (in prep.).
Therefore, in this Letter, we propose an alternative approach to sidestep the issue. Our new approach builds upon the Bayesian hierarchical likelihood construction method by directly seeding the priors of $x\Z1$ and $c$ with the inferred values from the SALT2 fitting algorithm \citep{Guy_2005, Guy_2007, Taylor_2021}. Specifically, we define the priors over the true values for each supernovae as 
\begin{equation}
    \prior\left[\left.(M\Ns{B}, x\Z 1, c) \Z i \right| H\right] = \mathcal{N}\left(M\Ns{B}|\bar{M}\Ns{B},\sigma_{M\Ns{B}}\right)\delta(x\Z 1-\hat{x}\Z {1,i})\,\delta(c-\hat{c}_{i}) \, ,
\end{equation}
where $\mathcal{N}\left(M\Ns{B}|\bar{M}\Ns{B},\sigma_{M\Ns{B}}\right)$ is a normal distribution with mean value $\bar{M}\Ns{B}$ and variance $\sigma_{M\Ns{B}}^2$, and $\delta$ is the Dirac delta distribution. Thus, the prior distribution in $M\Ns{B}$ is common to all the supernovae data, while the priors in $x\Z1$ and $c$ are supernovae specific. Therefore, our new approach sidesteps the problem of modelling the joint distribution, only requiring five parameters (a cosmological parameter, $\alpha$, $\beta$, $M\Ns B$, and $\sigma^2_{M\Ns B}$), by assuming that the SALT2 parameters represent the `true' parameters, i.e., the most probable values for both $x \Z 1$ and $c$ for this version of the SALT model.

Equivalently, given a single-shot inference for any physical quantity, the best guess for its true value is precisely the one inferred through the observational procedure. The assumption of being the most probable value introduces a caveat that it may, however, potentially overlook astrophysical systematics inherent in the SALT2 light curve procedure.

aOur approach here has essential differences from previous methodology \citep{Nielsen_2016, Dam_2017, Lane_2023}, and is not merely a change of priors. Earlier work assumed that all supernovae are drawn from ideal independent Gaussian distributions in stretch ($x\Z 1$) and in colour ($c$), with mean values and standard deviations derived from the cosmological fit. In contrast, this study does not assume any particular statistical distribution for $x\Z 1$ and $c$, nor do we assume these parameters follow the same ideal distribution across the supernova sample. Instead, $x\Z 1$ and $c$ are treated as fixed, with values provided by the SALT2 fit. \citet{Taylor_2021} show through simulations that SALT2 reliably recovers input supernova parameters. To compare this method with the previous one, we use the same dataset as \citet{Lane_2023}.

Therefore, by now following the same procedure as in \citet{Lane_2023}, we find the likelihood to be
\begin{align}
    \mathcal{L} &= (2 \pi)^{-3/2} \det\left[ D + \Sigma \Z d \right]^{-1/2} \exp\left[-\frac{1}{2}X^\intercal( D + \Sigma \Z {d})^{-1} X\right]\label{eq:likelihoodFull},
\end{align}
where the distributional error matrix ($D$) is the block-diagonal matrix with each block defined as ${\rm diag}\left( \sigma^2_{M_{\rm B}}, 0, 0\right)_i\,$, $\Sigma \Z d$ is the $3N \times 3N$ statistical and systematic covariance matrix given by \citet[Sec.~2]{Lane_2023}, and the residual vector $X$ is defined by 
\begin{eqnarray*}
X &:=& [\hat{m}_{\lower2pt\hbox{$\scriptstyle \rm B,1$}}^{*}-\mu _{\lower2pt\hbox{$\scriptstyle 1$}}-M_{\lower2pt\hbox{$\scriptstyle \rm B,1$}} + \alpha \hat{x}_{\lower2pt\hbox{$\scriptstyle 1,1$}}-\beta \hat{c}_{\lower2pt\hbox{$\scriptstyle 1$}}, 0, 0, ..., \nonumber \\
&&\hbox to3mm{\hfil } \hat{m}_{\lower2pt\hbox{$\scriptstyle \rm B, N$}}^{*}-\mu _{\lower2pt\hbox{$\scriptstyle N$}}-M_{\lower2pt\hbox{$\scriptstyle \rm B, N$}} + \alpha \hat{x}_{\lower2pt\hbox{$\scriptstyle 1, N$}}-\beta \hat{c}_{\lower2pt\hbox{$\scriptstyle N$}}, 0, 0]^\intercal \, . 
\end{eqnarray*}
Similarly to \citet{Dam_2017} and \citet{Lane_2023} we utilise a nested Bayesian sampler \texttt{PyMultiNest} \citep{Buchner_2014}, which interacts with the \texttt{MultiNest} \citep{Feroz_2008, Feroz_2009, Feroz_2019} code to compare the \lcdm\ and timescape models with a tolerance of $10^{-3}$ and $n\Ns{live} = 800$ for nine parameters. We choose the same priors as \citet[Table~B2 \& Sec.~3]{Lane_2023} summarised in \cref{tab:priors}. 

\begin{table}
\centering
\caption{Bayesian and frequentist priors on parameters used in the analysis. All priors are uniform on the respective intervals and, importantly, relatively broad for both models to ensure fair comparison.}
\begin{tabular}{c | c  } 
    \hline
     Parameter & Priors \\ 
    \hline
    \hline
    $f\Ns{v0}$ & [0.500,0.799] ($2 \sigma$ bound) \\ [0.8ex]
    $\Omega\Ns{M0}$ & [0.143,0.487] ($2 \sigma$ bound) \\ [0.8ex]
    $\alpha$ & [0,1] \\ [0.8ex] 
    $\beta$ & [0,7] \\ [0.8ex] 
    $x\Z 1$ & [-20,20] \\ [0.8ex] 
    $c$ & [-20,20] \\ [0.8ex] 
    $M\Ns{B}$ & [-20.3,18.3] \\ [0.8ex] 
    $\sigma\Z{M\Ns{B}}^2$ & $\log_{10}(\sigma)$ [-10,4] \\ [0.8ex] 
    \hline
\end{tabular}
\label{tab:priors}
\end{table}

Finally, in our analysis we reconstruct the $z \Ns {CMB}$ by applying a boost \citep{Fixsen_1996} to the Pantheon+ heliocentric redshifts, excluding peculiar velocity corrections. We then remove all supernovae with $z \Ns {CMB} \leq z \Ns {min}$ for varying redshift cuts $z\Ns {min}$ and fit the cosmological model to the remaining supernova events. This allows us to examine how the Bayes factor, cosmological parameters, and Tripp parameters vary across different redshift regimes.

\section{Results}\label{sec:results}

\begin{figure}
    \includegraphics[width=\columnwidth]{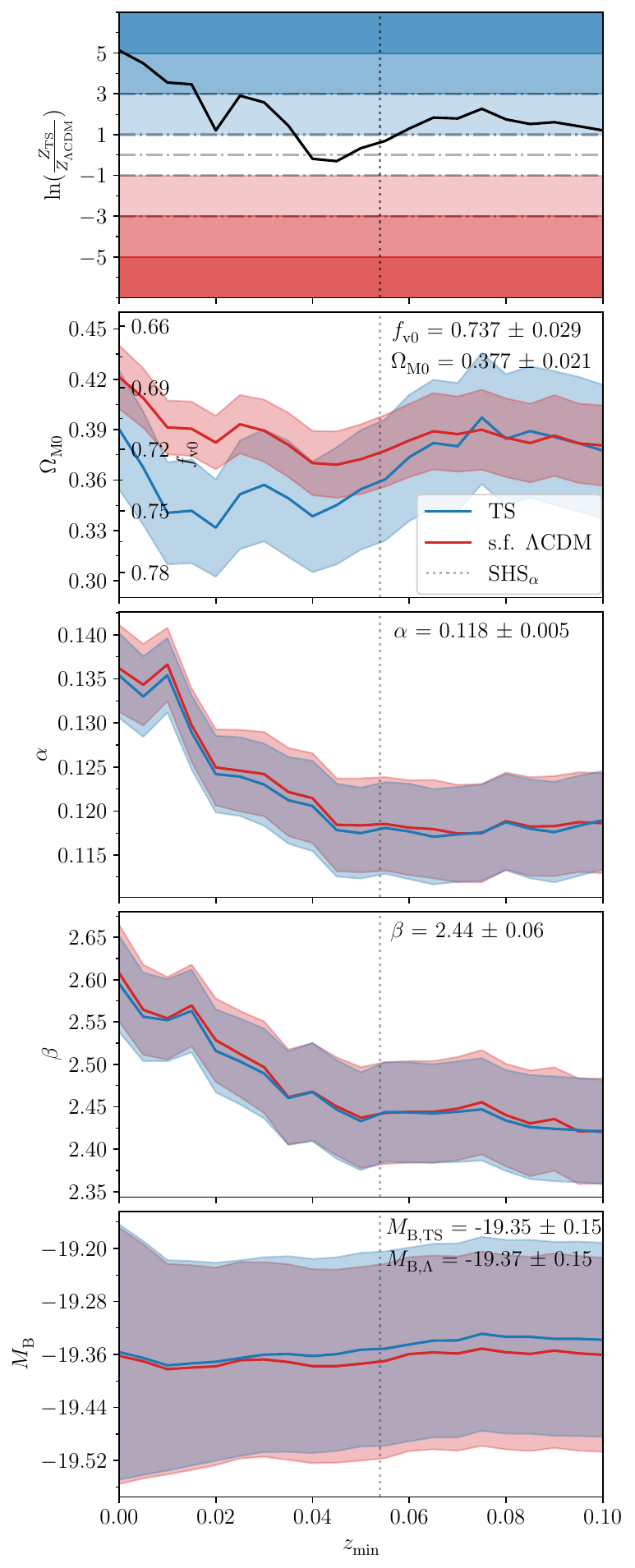}
    \caption{The Bayes factors and Bayesian Maximum Likelihood Estimate (MLE) parameters for the fitting parameters across different redshift cuts, with Bayes factor uncertainties too small to display in the plot. The top plot shows the Bayes factors, here the blue ($\ln B > 1$) section favours timescape, the white section favours neither hypothesis and the red ($\ln B < -1$) section favours \LCDM. The following plots show the various MLE parameter estimates, with values beyond SHS$_\alpha = 0.054^{+0.007}_{-0.012}$ indicated by the dashed vertical line.}
    \label{fig:results}
\end{figure}

Results for the Bayes factor, cosmological and light-curve parameters are shown in \cref{fig:results}.

The Bayesian comparisons are best understood by splitting the minimum redshift cutoff used into three regimes: (i) for $0< \zmin < 0.023$ we find very strong to strong evidence on the Jeffrey's scale \citep{Kass_1995} in favour of timescape over \LCDM; (ii) for $0.023\leq \zmin < 0.054$ we enter the \textit{calibration regime},\footnote{This is the regime beyond which average homogeneity and isotropy are assumed to apply to all observations. \citet{Hogg_2005, Scrimgeour_2012} take this range as $\gsim70$--$120\hM$ (corresponding to a redshift range of approximately 0.023 -- 0.04).} finding moderate to no significant preference for timescape; (iii) for $\zmin \geq 0.075$, beyond any measure of a SHS\footnote{The \citet{Lane_2023} value $\zmin=0.075$ is larger than other estimates and thus gives a robust upper bound for the SHS.}, we find exclusively moderate preference for the timescape cosmology. Notably, the log-evidence, $\ln Z$, values found here for both models are $\sim 10-100\times$ greater compared to the previous analysis by \citet{Lane_2023}.

Since timescape's uniform quasilocal Hubble expansion condition holds down to scales $\goesas4\,$--$\,15\,$Mpc, as we decrease $\zmin$ an increase in the Bayesian evidence favouring timescape is expected if the model accurately captures the average cosmic expansion deep in the nonlinear regime of structure formation. Beyond the SHS, \LCDM\ of course provides an excellent description of our Universe. However, the evidence in favour of timescape remains small but modest ($\ln B > 1$) at the highest redshift cuts, $\zmin \geq 0.075$, pointing to the ability of the model to describe the Universe's expansion history on scales greater than the SHS. This moderate evidence ($\ln B > 1$) can be interpreted as resulting from the integrated effects across the redshift range $\zmin<z<2.26$, reflecting the $1$--$3$\% variations in the expansion history between timescape and \LCDM.

\begin{figure}
    \includegraphics[width=\columnwidth]{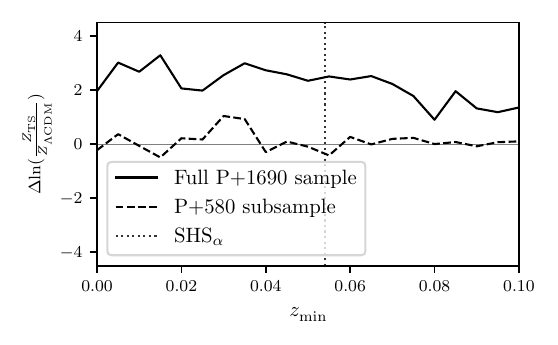}
    \caption{The difference in the Bayes factors for the full P+1690 sample and the P+580 subsample between \laneetal\ and our results. For the subsample, the results from the new analysis presented here align very well with the results by \laneetal, while for the full sample the new analysis greatly increases the preference in favour of timescape.}
    \label{fig:results580}
\end{figure}

In comparing two models with different assumptions in the nonlinear regime, the redshift distribution of the data becomes particularly important. For example, \laneetal\ found consistent weak preference in favour of timescape using the P+580 subsample in which data from the full sample is truncated at high and low redshifts. While the evidence for the P+1690 sample changes significantly of order $\ln B \sim 1.5\,$--$\,2.5$  in our revised analysis, the P+580 subsample result remains consistent (\cref{fig:results580}). The discrepancy between the results of the full dataset and the subsample suggest the need for further analysis on how the redshift distribution of supernovae, and the probed redshift range impact evidence for cosmological models. The uncertainty in the Bayes factor,  $\sim0.014$, is so small that it does not influence the Jeffrey's scale classifications or the conclusions drawn.

The Bayes factors and Bayesian Maximum Likelihood Estimate (MLE) parameters for different redshift cuts are shown. The top panel shows Bayes factors with blue indicating preference for timescape, red for \LCDM, and white for neither. The subsequent plots show MLE parameter estimates, with values beyond the scale of the statistical homogeneity (SHS) marked by the dashed vertical line.

\begin{figure}
    \includegraphics[width=\columnwidth]{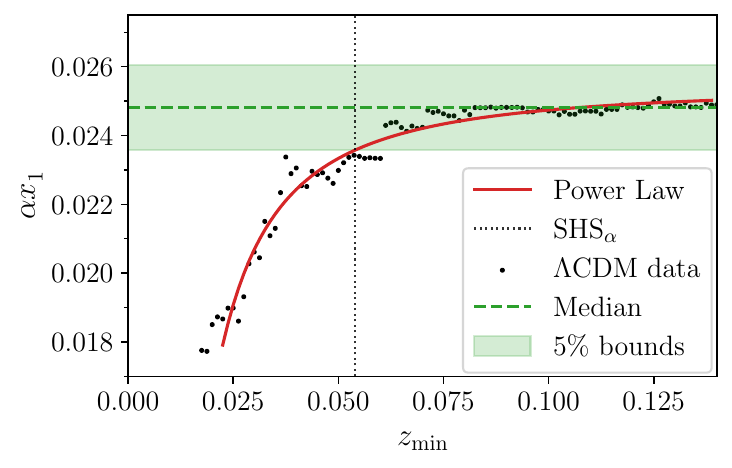}
    \caption{The convergence of the $\alpha x\Z 1$ light-curve parameter for the \lcdm\ model across various redshift cuts, where $x \Z 1$ is the median value from the distribution. A power--law model has been fit to the data, and the green shaded band represents within 5\% of the median value within the range $0.1 \leq \zmin < 0.14$ indicating when the model converges. The vertical dotted line represents the SHS$_\alpha$ found at $\zmin = 0.054^{+0.007}_{-0.012}$. The power-law uncertainty is smaller than the plotted line.}
    \label{fig:shs}
\end{figure}

\citet{Lane_2023} introduced an additional empirical data-driven notion of statistical homogeneity, defining SHS$_\alpha$ from a power--law fitted to the $\alpha x \Z 1$ degenerate parameter. The analogous SHS$_\beta$ defined from $\beta c$ does not yield a true convergence for the analysis by \citet{Lane_2023}, nor for this analysis, due to Malmquist bias not being accounted for. While the SHS$_\beta$ appears to converge below $\zmin \approx 0.12$, this is not the case for higher redshift cuts. For the reanalysis presented in \cref{fig:shs} we find SHS$_\alpha = 0.054_{-0.012}^{+0.007}$, which is 1.2$\sigma$ greater than the maximum value of the SHS gathered from the two-point galaxy correlation function \citep{Hogg_2005, Scrimgeour_2012} and somewhat lower but within 2.3$\sigma$ of the result \citet{Lane_2023} determined. The differences with respect to the analysis in \citet{Lane_2023} derive from the lifting of the Gaussian assumption of the underlying distributions of $x\Z1$ and $c$.

The Bayesian analysis can be used to find the maximum likelihood estimate (MLE) of the parameters, including the single free cosmological parameter. For $\zmin$ cuts beyond the SHS$_\alpha$, ($\zmin \geq 0.055$), for \LCDM\ we find $\Omega\Ns{M0}= 0.377 \pm 0.021$, within 1.2$\sigma$ of the value found from the DES5yr release \citep{DES_2024}, and just outside of 2$\sigma$ of \pplus\ \citep{Brout_2022_cosmo}.\footnote{The $\alpha$ and $\beta$ values reported by \citet{Scolnic_2022} and \laneetal\ are derived at various stages of the cosmological fitting pipeline, and are influenced by the specific subsample used \citep{Lane_2023}. Any slight differences in cosmological parameters can be attributed to these methodological variations and to the omission of cosmology dependent bias corrections.}

In the case of timescape we find a void fraction of, $f\Ns{v0} = 0.737 \pm 0.029$, within 2$\sigma$ of the \citet{Camilleri_2024} DES5yr value. Significantly, our $f\Ns{v0}$ value is also within 2$\sigma$ of independent values predicted from the Planck CMB power spectrum, $f\Ns{v0} = 0.695^{+0.041}_{-0.051}$ \citep{Duley_2013}; and well within 1$\sigma$ of strong gravitational lensing distance ratios, $f\Ns{v0} = 0.736 \pm 0.099$, \citep{HarveyHawes_2024}. We also find the evolution of the Tripp constants, $\alpha$, $\beta$, and $M\Ns{B}$ with varying $z\Ns{min}$ cuts following \citet{Dam_2017, Lane_2023}. To avoid the underlying degeneracy between $\Hn$ and $M\Ns{B}$, we fix $\Hn$ for both models as a nuisance parameter.\footnote{The relative contributions of Hubble constant uncertainty and absolute magnitude uncertainty, respectively $\delta\Z{\Hn}$ and $\delta\Z{M\Ns{B}}$, propagate according to
  $ \sigma\Z{M\Ns{B}} = \Bigl[\delta\Z{M\Ns{B}}^2 + \Bigl( \frac{5}{\Hn \ln{10}} \Bigr)^2 \delta\Z{\Hn}^2\Bigr]^{1/2}\,.$
This makes the two contributions impossible to unravel and explains the larger uncertainty, $\sigma\Z{M\Ns{B}}$,
relative to uncertainties in other parameters from the fitting (see \cref{fig:results}).} Moreover, although the values for the individual parameters differ between the two statistical methods, the Tripp distance modulus, $\mu$, changes on average by only $| \Delta (- M \Ns{B} + \alpha x\Z1 - \beta c) |_{\rm \Lambda CDM} = 0.030 \pm 0.019$ for redshift cuts beyond $z \Ns {min}=0.054$. This variation is observed when comparing the median values of $x\Z 1$ and $c$ in the Tripp methodology, to the general Gaussian distribution fit values.

The change in $\mu$ between this work and \citet{Lane_2023} is thus not statistically significant in this regime. However, it is expected that differences in the prior distribution cause differences in the fitted parameters. This behaviour will be investigated further for supernovae statistics built on skewed, non-Gaussian distributions in future work (in prep.).

In the Beams with Bias Correction method \citep{Kessler_2017} a galaxy host correction is introduced with an additional parameter, $\gamma$, defined by the mass-step
\begin{equation}
\gamma G\Ns{Host} = \delta\Ns{G. Host} = 
    \begin{cases}
        \gamma/2 & \text{if } {\rm M}\Ns{Galaxy} \geq 10^{10} {\rm M}\Z{\odot}\\
        -\gamma/2 & \text{if } {\rm M}\Ns{Galaxy} < 10^{10} {\rm M}\Z{\odot}
    \end{cases}
\end{equation}
We examined including this term but found that it does not affect the Bayes factor conclusions, with an average
offset of $| \Delta \ln B | = 0.077$ compared to the uncorrected value. Furthermore, the statistical cost of introducing additional free parameters can be assessed by the relative Bayesian Information Criterion (BIC) statistic \citep{Schwarz_1978, Kass_1995} ${\rm BIC} = k\ln N - 2 \ln Z\,, $ for $k$ free parameters, a sample size, $N$, and likelihood $Z$. We find that independent of cosmology the model with a mass step is strongly disfavoured relative to the uncorrected Tripp model, with $\Delta {\rm BIC} = -7.90$ at $\zmin = 0$. Furthermore, there is no significant change in the value of the cosmological parameter, with $\Delta \Omega\Ns{M0} \approx 0.0016$, which is well within the 1$\sigma$ range of our statistical and systematic uncertainties. Thus our final results are stated without galaxy host corrections.\footnote{A further reason for not including galaxy-host corrections is the observation that the $\gamma$ parameter exhibits inconsistent behaviour across different redshift cuts for a simple mass-step function. This inconsistency most likely arises from the heterogeneous subsamples of low-$z$ data, as $\gamma$ is well-constrained in a more statistically homogeneous sample. It is possible, but less likely, that these fluctuations arise from other astrophysical factors explored by the DES5yr team and recent studies \citep{Dixon_2024}, but these are beyond the scope of this Letter.}

\section{Discussion and Conclusions}
We performed a new Bayesian statistical analysis on the \pplus\ supernovae data set, accounting for the non-Gaussian $x\Z 1$ and $c$ features of the supernovae parameter distributions. The Bayesian evidence yields very strong to strong evidence for the timescape model in the low redshift regime. This late-universe result could be expected, as the timescape models accounts for non-kinematic differential expansion on scales $z\lsim0.03$ where the local inhomogeneous structure of our nearby cosmic web most impacts measurements.
On the other hand, for samples strongly weighted by \sne\ in the calibration regime of the \LCDM\ model ($z\Ns{CMB}\approx0.04$) there is no significant preference either way, the two models being statistically equivalent. With a restriction to higher redshifts, well beyond any scale of statistical homogeneity generally accepted \citep{Lane_2023}, Bayesian evidence is driven once again in favour of timescape.

Our new analysis makes fewer assumptions about any particular statistical distribution of the data. Specifically, the likelihood function is constructed directly from the $x\Z 1$ and $c$ values obtained using the SALT2 algorithm -- values employed in most \sne\ analyses. The empirical \sne\ data obtained via the cosmology independent SALT2 fit strongly favours the timescape model over \LCDM.

Any astrophysical or environmental biases would likely impact both cosmological models. Thus the strong preference for timescape would require an extremely subtle combination of such biases for this to be its prime cause.
The largest systematic error in the \pplus\ analysis is the standardisation of the heterogeneous mix of low-$z$ sample light curves \citep{DES_2024, Lane_2023}.  Future improvements with the new DES5yr sample \citep{DES_2024} will allow for a more homogeneous and careful selection of the low-$z$ sample. However, in this Letter we concentrate on the impact of the new statistical method on cosmological model selection, and therefore we use the same data as \laneetal. 

Since timescape has the same number of free parameters as \lcdm, Bayesian evidence offers the best comparison. To expand our results to include other popular FLRW-type alternative cosmological models, which contain more parameters, e.g., $w$CDM, we determine the BIC statistic \citep{Schwarz_1978, Kass_1995} for fair model comparison. For the full sample, we find that relative to timescape \LCDM\ models with FLRW curvature are very strongly disfavoured with $\Delta{\rm BIC} = -13.39$, while $w$CDM is also very strongly disfavoured with $\Delta{\rm BIC} = -11.70$.

The results presented in this Letter indicate that the timescape cosmology is not only a viable contender to the \LCDM\ framework, but may also provide new insights to the astrophysics of modelling \sne. 
Timescape's non-FLRW average evolution reveals degeneracies between cosmological parameters and empirical \sne\ model parameters that were already partly uncovered in earlier work \citep{Dam_2017} but which are striking with \pplus, as shown by \citet{Lane_2023} and the present Letter. 

Regardless of what model cosmology is to be the standard in future, exploring more than one model is important. Indeed, the timescape framework is consistent with new analysis of void statistics in numerical relativity simulations using the full Einstein equations \citep{Williams_2024}. These are consistent with an emerging kinetic spatial curvature of voids on small scales. Much remains to be done in calibrating the dark matter fraction, primordial sound speed and the BAO scale. However, new results are likely to provide a robust framework for this \citep{Galoppo_2024a,Galoppo_2024b}.

Our results imply profound consequences for cosmology and astrophysics. Indeed, a net preference for the timescape cosmology over the standard FLRW cosmologies may point to a need for revision of the foundations of theoretical cosmology, both ontologically and epistemologically, to better understand inhomogeneities and their backreaction on the average evolution of the Universe.

\section*{Acknowledgements}
DLW, RRH and ZGL are supported by the Marsden Fund administered by the Royal Society of New Zealand, Te Apārangi under grants M1271 and M1255. RRH is also supported by Rutherford Foundation Postdoctoral Fellowship RFT-UOC2203-PD. We are indebted to the anonymous referee of \citet{Lane_2023} for suggesting the analysis framework. We thank all members of the University of Canterbury Gravity and Cosmology and Astrophysics groups for stimulating discussions, particularly: John Forbes, Christopher Harvey-Hawes, Morag Hills, Emma Johnson, Shreyas Tiruvaskar, Michael Williams, and Manon van Zyl. Finally, we wish to thank Elena Moltchanova for her precious insights into the statistical methods employed.

\section*{Data Availability}

A complete set of the codes and details used for our analysis and how to use them can
be found at \citet{Seifert_2023}, and the covariance and input files are made available at \citet{Lane_2024}.


\bibliographystyle{mnras}
\bibliography{references}

\begin{thebibliography}{}
\makeatletter
\relax
\def\mn@urlcharsother{\let\do\@makeother \do\$\do\&\do\#\do\^\do\_\do\%\do\~}
\def\mn@doi{\begingroup\mn@urlcharsother \@ifnextchar [ {\mn@doi@} {\mn@doi@[]}}
\def\mn@doi@[#1]#2{\def\@tempa{#1}\ifx\@tempa\@empty \href {http://dx.doi.org/#2} {doi:#2}\else \href {http://dx.doi.org/#2} {#1}\fi \endgroup}
\def\mn@eprint#1#2{\mn@eprint@#1:#2::\@nil}
\def\mn@eprint@arXiv#1{\href {http://arxiv.org/abs/#1} {{\tt arXiv:#1}}}
\def\mn@eprint@dblp#1{\href {http://dblp.uni-trier.de/rec/bibtex/#1.xml} {dblp:#1}}
\def\mn@eprint@#1:#2:#3:#4\@nil{\def\@tempa {#1}\def\@tempb {#2}\def\@tempc {#3}\ifx \@tempc \@empty \let \@tempc \@tempb \let \@tempb \@tempa \fi \ifx \@tempb \@empty \def\@tempb {arXiv}\fi \@ifundefined {mn@eprint@\@tempb}{\@tempb:\@tempc}{\expandafter \expandafter \csname mn@eprint@\@tempb\endcsname \expandafter{\@tempc}}}

\bibitem[\protect\citeauthoryear{{Abbott} et~al.,}{{Abbott} et~al.}{2024}]{DES_2024}
{Abbott} T.~M.~C.,  et~al., 2024, \mn@doi [\apjl] {10.3847/2041-8213/ad6f9f}, \href {https://ui.adsabs.harvard.edu/abs/2024ApJ...973L..14A} {973, L14}

\bibitem[\protect\citeauthoryear{{Adame} et~al.,}{{Adame} et~al.}{2024}]{DESI_2024}
{Adame} A.~G.,  et~al., 2024,  (\mn@eprint {arXiv} {2404.03002})

\bibitem[\protect\citeauthoryear{{Aluri} et~al.,}{{Aluri} et~al.}{2023}]{Aluri_2022}
{Aluri} P.~K.,  et~al., 2023, \mn@doi [Classical and Quantum Gravity] {10.1088/1361-6382/acbefc}, \href {https://ui.adsabs.harvard.edu/abs/2023CQGra..40i4001K} {40, 094001}

\bibitem[\protect\citeauthoryear{{Betoule} et~al.,}{{Betoule} et~al.}{2014}]{Betoule_2014}
{Betoule} M.,  et~al., 2014, \mn@doi [\aap] {10.1051/0004-6361/201423413}, \href {https://ui.adsabs.harvard.edu/abs/2014A&A...568A..22B} {568, A22}

\bibitem[\protect\citeauthoryear{{Brout} et~al.,}{{Brout} et~al.}{2022a}]{Brout_2022_cosmo}
{Brout} D.,  et~al., 2022a, \mn@doi [\apj] {10.3847/1538-4357/ac8e04}, \href {https://ui.adsabs.harvard.edu/abs/2022ApJ...938..110B} {938, 110}

\bibitem[\protect\citeauthoryear{{Brout} et~al.,}{{Brout} et~al.}{2022b}]{Brout_2022_cal}
{Brout} D.,  et~al., 2022b, \mn@doi [\apj] {10.3847/1538-4357/ac8bcc}, \href {https://ui.adsabs.harvard.edu/abs/2022ApJ...938..111B} {938, 111}

\bibitem[\protect\citeauthoryear{{Buchert}}{{Buchert}}{2000}]{Buchert_2000}
{Buchert} T.,  2000, \mn@doi [General Relativity and Gravitation] {10.1023/A:1001800617177}, \href {https://ui.adsabs.harvard.edu/abs/2000GReGr..32..105B} {32, 105}

\bibitem[\protect\citeauthoryear{{Buchert}}{{Buchert}}{2001}]{Buchert_2001}
{Buchert} T.,  2001, \mn@doi [General Relativity and Gravitation] {10.1023/A:1012061725841}, \href {https://ui.adsabs.harvard.edu/abs/2001GReGr..33.1381B} {33, 1381}

\bibitem[\protect\citeauthoryear{{Buchert}, {Mourier}  \& {Roy}}{{Buchert} et~al.}{2020}]{Buchert_2020}
{Buchert} T.,  {Mourier} P.,   {Roy} X.,  2020, \mn@doi [Gen.\ Relativ.\ Gravitation] {10.1007/s10714-020-02670-6}, \href {https://ui.adsabs.harvard.edu/abs/2020GReGr..52...27B} {52, 27}

\bibitem[\protect\citeauthoryear{{Buchner} et~al.,}{{Buchner} et~al.}{2014}]{Buchner_2014}
{Buchner} J.,  et~al., 2014, \mn@doi [\aap] {10.1051/0004-6361/201322971}, \href {https://ui.adsabs.harvard.edu/abs/2014A&A...564A.125B} {564, A125}

\bibitem[\protect\citeauthoryear{{Camilleri} et~al.,}{{Camilleri} et~al.}{2024}]{Camilleri_2024}
{Camilleri} R.,  et~al., 2024, \mn@doi [\mnras] {10.1093/mnras/stae1988}, \href {https://ui.adsabs.harvard.edu/abs/2024MNRAS.533.2615C} {533, 2615}

\bibitem[\protect\citeauthoryear{{Carr}, {Davis}, {Scolnic}, {Said}, {Brout}, {Peterson}  \& {Kessler}}{{Carr} et~al.}{2022}]{Carr_2022}
{Carr} A.,  {Davis} T.~M.,  {Scolnic} D.,  {Said} K.,  {Brout} D.,  {Peterson} E.~R.,   {Kessler} R.,  2022, \mn@doi [\pasa] {10.1017/pasa.2022.41}, \href {https://ui.adsabs.harvard.edu/abs/2022PASA...39...46C} {39, e046}

\bibitem[\protect\citeauthoryear{{Dam}, {Heinesen}  \& {Wiltshire}}{{Dam} et~al.}{2017}]{Dam_2017}
{Dam} L.~H.,  {Heinesen} A.,   {Wiltshire} D.~L.,  2017, \mn@doi [\mnras] {10.1093/mnras/stx1858}, \href {https://ui.adsabs.harvard.edu/abs/2017MNRAS.472..835D} {472, 835}

\bibitem[\protect\citeauthoryear{{Di Valentino} et~al.,}{{Di Valentino} et~al.}{2021}]{Di_Valentino_2021}
{Di Valentino} E.,  et~al., 2021, \mn@doi [Classical and Quantum Gravity] {10.1088/1361-6382/ac086d}, \href {https://ui.adsabs.harvard.edu/abs/2021CQGra..38o3001D} {38, 153001}

\bibitem[\protect\citeauthoryear{{Dixon} et~al.,}{{Dixon} et~al.}{2024}]{Dixon_2024}
{Dixon} M.,  et~al., 2024,  (\mn@eprint {arXiv} {2408.01001})

\bibitem[\protect\citeauthoryear{{Duley}, {Nazer}  \& {Wiltshire}}{{Duley} et~al.}{2013}]{Duley_2013}
{Duley} J. A.~G.,  {Nazer} M.~A.,   {Wiltshire} D.~L.,  2013, \mn@doi [Classical and Quantum Gravity] {10.1088/0264-9381/30/17/175006}, \href {https://ui.adsabs.harvard.edu/abs/2013CQGra..30q5006D} {30, 175006}

\bibitem[\protect\citeauthoryear{{Feroz} \& {Hobson}}{{Feroz} \& {Hobson}}{2008}]{Feroz_2008}
{Feroz} F.,  {Hobson} M.~P.,  2008, \mn@doi [\mnras] {10.1111/j.1365-2966.2007.12353.x}, \href {https://ui.adsabs.harvard.edu/abs/2008MNRAS.384..449F} {384, 449}

\bibitem[\protect\citeauthoryear{{Feroz}, {Hobson}  \& {Bridges}}{{Feroz} et~al.}{2009}]{Feroz_2009}
{Feroz} F.,  {Hobson} M.~P.,   {Bridges} M.,  2009, \mn@doi [\mnras] {10.1111/j.1365-2966.2009.14548.x}, \href {https://ui.adsabs.harvard.edu/abs/2009MNRAS.398.1601F} {398, 1601}

\bibitem[\protect\citeauthoryear{{Feroz}, {Hobson}, {Cameron}  \& {Pettitt}}{{Feroz} et~al.}{2019}]{Feroz_2019}
{Feroz} F.,  {Hobson} M.~P.,  {Cameron} E.,   {Pettitt} A.~N.,  2019, \mn@doi [The Open Journal of Astrophysics] {10.21105/astro.1306.2144}, \href {https://ui.adsabs.harvard.edu/abs/2019OJAp....2E..10F} {2, 1}

\bibitem[\protect\citeauthoryear{{Fixsen}, {Cheng}, {Gales}, {Mather}, {Shafer}  \& {Wright}}{{Fixsen} et~al.}{1996}]{Fixsen_1996}
{Fixsen} D.~J.,  {Cheng} E.~S.,  {Gales} J.~M.,  {Mather} J.~C.,  {Shafer} R.~A.,   {Wright} E.~L.,  1996, \mn@doi [\apj] {10.1086/178173}, \href {https://ui.adsabs.harvard.edu/abs/1996ApJ...473..576F} {473, 576}

\bibitem[\protect\citeauthoryear{{Galoppo} \& {Wiltshire}}{{Galoppo} \& {Wiltshire}}{2024}]{Galoppo_2024a}
{Galoppo} M.,  {Wiltshire} D.~L.,  2024,  (\mn@eprint {arXiv} {2406.14157})

\bibitem[\protect\citeauthoryear{{Galoppo}, {Re}  \& {Wiltshire}}{{Galoppo} et~al.}{2024}]{Galoppo_2024b}
{Galoppo} M.,  {Re} F.,   {Wiltshire} D.~L.,  2024,  (\mn@eprint {arXiv} {2408.00385})

\bibitem[\protect\citeauthoryear{{Guy}, {Astier}, {Nobili}, {Regnault}  \& {Pain}}{{Guy} et~al.}{2005}]{Guy_2005}
{Guy} J.,  {Astier} P.,  {Nobili} S.,  {Regnault} N.,   {Pain} R.,  2005, \mn@doi [\aap] {10.1051/0004-6361:20053025}, \href {https://ui.adsabs.harvard.edu/abs/2005A&A...443..781G} {443, 781}

\bibitem[\protect\citeauthoryear{{Guy} et~al.,}{{Guy} et~al.}{2007}]{Guy_2007}
{Guy} J.,  et~al., 2007, \mn@doi [\aap] {10.1051/0004-6361:20066930}, \href {https://ui.adsabs.harvard.edu/abs/2007A&A...466...11G} {466, 11}

\bibitem[\protect\citeauthoryear{{Harvey-Hawes} \& {Wiltshire}}{{Harvey-Hawes} \& {Wiltshire}}{2024}]{HarveyHawes_2024}
{Harvey-Hawes} C.,  {Wiltshire} D.~L.,  2024, \mn@doi [\mnras] {10.1093/mnras/stae2306}, \href {https://ui.adsabs.harvard.edu/abs/2024MNRAS.534.3364H} {534, 3364}

\bibitem[\protect\citeauthoryear{{Heinesen}, {Blake}, {Li}  \& {Wiltshire}}{{Heinesen} et~al.}{2019}]{Heinesen_2019}
{Heinesen} A.,  {Blake} C.,  {Li} Y.-Z.,   {Wiltshire} D.~L.,  2019, \mn@doi [\jcap] {10.1088/1475-7516/2019/03/003}, \href {https://ui.adsabs.harvard.edu/abs/2019JCAP...03..003H} {{\protect\rm03}, 003}

\bibitem[\protect\citeauthoryear{{Hinton} et~al.,}{{Hinton} et~al.}{2019}]{Hinton_2019}
{Hinton} S.~R.,  et~al., 2019, \mn@doi [\apj] {10.3847/1538-4357/ab13a3}, \href {https://ui.adsabs.harvard.edu/abs/2019ApJ...876...15H} {876, 15}

\bibitem[\protect\citeauthoryear{{Hogg}, {Eisenstein}, {Blanton}, {Bahcall}, {Brinkmann}, {Gunn}  \& {Schneider}}{{Hogg} et~al.}{2005}]{Hogg_2005}
{Hogg} D.~W.,  {Eisenstein} D.~J.,  {Blanton} M.~R.,  {Bahcall} N.~A.,  {Brinkmann} J.,  {Gunn} J.~E.,   {Schneider} D.~P.,  2005, \mn@doi [\apj] {10.1086/429084}, \href {https://ui.adsabs.harvard.edu/abs/2005ApJ...624...54H} {624, 54}

\bibitem[\protect\citeauthoryear{Kass \& Raftery}{Kass \& Raftery}{1995}]{Kass_1995}
Kass R.~E.,  Raftery A.~E.,  1995, \mn@doi [Journal of the American Statistical Association] {10.1080/01621459.1995.10476572}, 90, 773

\bibitem[\protect\citeauthoryear{{Kenworthy} et~al.,}{{Kenworthy} et~al.}{2021}]{Kenworthy_2021}
{Kenworthy} W.~D.,  et~al., 2021, \mn@doi [\apj] {10.3847/1538-4357/ac30d8}, \href {https://ui.adsabs.harvard.edu/abs/2021ApJ...923..265K} {923, 265}

\bibitem[\protect\citeauthoryear{{Kessler} \& {Scolnic}}{{Kessler} \& {Scolnic}}{2017}]{Kessler_2017}
{Kessler} R.,  {Scolnic} D.,  2017, \mn@doi [\apj] {10.3847/1538-4357/836/1/56}, \href {https://ui.adsabs.harvard.edu/abs/2017ApJ...836...56K} {836, 56}

\bibitem[\protect\citeauthoryear{{Kessler} et~al.,}{{Kessler} et~al.}{2009}]{Kessler_2009_Sloan}
{Kessler} R.,  et~al., 2009, \mn@doi [\apjs] {10.1088/0067-0049/185/1/32}, \href {https://ui.adsabs.harvard.edu/abs/2009ApJS..185...32K} {185, 32}

\bibitem[\protect\citeauthoryear{Lane \& Seifert}{Lane \& Seifert}{2024}]{Lane_2024}
Lane Z.~G.,  Seifert A.,  2024, P+1690 Covariance Matrix, \url {https://doi.org/10.5281/zenodo.12729746}

\bibitem[\protect\citeauthoryear{{Lane}, {Seifert}, {Ridden-Harper}  \& {Wiltshire}}{{Lane} et~al.}{2025}]{Lane_2023}
{Lane} Z.~G.,  {Seifert} A.,  {Ridden-Harper} R.,   {Wiltshire} D.~L.,  2025, \mn@doi [\mnras] {10.1093/mnras/stae2437}, \href {https://ui.adsabs.harvard.edu/abs/2024MNRAS.tmp.2463L} {536, 1752}

\bibitem[\protect\citeauthoryear{{March}, {Trotta}, {Berkes}, {Starkman}  \& {Vaudrevange}}{{March} et~al.}{2011}]{March_2011}
{March} M.~C.,  {Trotta} R.,  {Berkes} P.,  {Starkman} G.~D.,   {Vaudrevange} P.~M.,  2011, \mn@doi [\mnras] {10.1111/j.1365-2966.2011.19584.x}, \href {https://ui.adsabs.harvard.edu/abs/2011MNRAS.418.2308M} {418, 2308}

\bibitem[\protect\citeauthoryear{McKay}{McKay}{2016}]{McKay_thesis_2016}
McKay J.~H.,  2016, \protect{MSc} thesis, University of Canterbury, \mn@doi{10.26021/6882}

\bibitem[\protect\citeauthoryear{{Nielsen}, {Guffanti}  \& {Sarkar}}{{Nielsen} et~al.}{2016}]{Nielsen_2016}
{Nielsen} J.~T.,  {Guffanti} A.,   {Sarkar} S.,  2016, \mn@doi [Scientific Reports] {10.1038/srep35596}, \href {https://ui.adsabs.harvard.edu/abs/2016NatSR...635596N} {6, 35596}

\bibitem[\protect\citeauthoryear{{Peebles}}{{Peebles}}{2022}]{Peebles_2022}
{Peebles} P.~J.~E.,  2022, \mn@doi [Annals of Physics] {10.1016/j.aop.2022.169159}, \href {https://ui.adsabs.harvard.edu/abs/2022AnPhy.44769159P} {447, 169159}

\bibitem[\protect\citeauthoryear{{Perlmutter} et~al.,}{{Perlmutter} et~al.}{1999}]{Perlmutter_1999}
{Perlmutter} S.,  et~al., 1999, \mn@doi [\apj] {10.1086/307221}, \href {https://ui.adsabs.harvard.edu/abs/1999ApJ...517..565P} {517, 565}

\bibitem[\protect\citeauthoryear{{Riess} et~al.,}{{Riess} et~al.}{1998}]{Riess_1998}
{Riess} A.~G.,  et~al., 1998, \mn@doi [\aj] {10.1086/300499}, \href {https://ui.adsabs.harvard.edu/abs/1998AJ....116.1009R} {116, 1009}

\bibitem[\protect\citeauthoryear{{Schwarz}}{{Schwarz}}{1978}]{Schwarz_1978}
{Schwarz} G.,  1978, Annals of Statistics, \href {https://ui.adsabs.harvard.edu/abs/1978AnSta...6..461S} {6, 461}

\bibitem[\protect\citeauthoryear{{Scolnic} et~al.,}{{Scolnic} et~al.}{2022}]{Scolnic_2022}
{Scolnic} D.,  et~al., 2022, \mn@doi [\apj] {10.3847/1538-4357/ac8b7a}, \href {https://ui.adsabs.harvard.edu/abs/2022ApJ...938..113S} {938, 113}

\bibitem[\protect\citeauthoryear{{Scrimgeour} et~al.,}{{Scrimgeour} et~al.}{2012}]{Scrimgeour_2012}
{Scrimgeour} M.~I.,  et~al., 2012, \mn@doi [\mnras] {10.1111/j.1365-2966.2012.21402.x}, \href {https://ui.adsabs.harvard.edu/abs/2012MNRAS.425..116S} {425, 116}

\bibitem[\protect\citeauthoryear{Seifert \& Lane}{Seifert \& Lane}{2023}]{Seifert_2023}
Seifert A.,  Lane Z.~G.,  2023, Code for Supernova Analysis with Pantheon+ by Lane et al. 2024, \url {https://github.com/antosft/SNe-PantheonPlus-Analysis}

\bibitem[\protect\citeauthoryear{{Taylor}, {Lidman}, {Tucker}, {Brout}, {Hinton}  \& {Kessler}}{{Taylor} et~al.}{2021}]{Taylor_2021}
{Taylor} G.,  {Lidman} C.,  {Tucker} B.~E.,  {Brout} D.,  {Hinton} S.~R.,   {Kessler} R.,  2021, \mn@doi [\mnras] {10.1093/mnras/stab962}, \href {https://ui.adsabs.harvard.edu/abs/2021MNRAS.504.4111T} {504, 4111}

\bibitem[\protect\citeauthoryear{Tripp}{Tripp}{1998}]{Tripp_1998}
Tripp R.,  1998, \aap, 331, 815

\bibitem[\protect\citeauthoryear{{Williams}, {Macpherson}, {Wiltshire}  \& {Stevens}}{{Williams} et~al.}{2024}]{Williams_2024}
{Williams} M.~J.,  {Macpherson} H.~J.,  {Wiltshire} D.~L.,   {Stevens} C.,  2024,  (\mn@eprint {arXiv} {2403.11997})

\bibitem[\protect\citeauthoryear{{Wiltshire}}{{Wiltshire}}{2007a}]{Wiltshire_2007_clocks}
{Wiltshire} D.~L.,  2007a, \mn@doi [New Journal of Physics] {10.1088/1367-2630/9/10/377}, \href {https://ui.adsabs.harvard.edu/abs/2007NJPh....9..377W} {9, 377}

\bibitem[\protect\citeauthoryear{{Wiltshire}}{{Wiltshire}}{2007b}]{Wiltshire_2007_sol}
{Wiltshire} D.~L.,  2007b, \mn@doi [\prl] {10.1103/PhysRevLett.99.251101}, \href {https://ui.adsabs.harvard.edu/abs/2007PhRvL..99y1101W} {99, 251101}

\bibitem[\protect\citeauthoryear{{Wiltshire}}{{Wiltshire}}{2008}]{Wiltshire_2008}
{Wiltshire} D.~L.,  2008, \mn@doi [\prd] {10.1103/PhysRevD.78.084032}, \href {https://ui.adsabs.harvard.edu/abs/2008PhRvD..78h4032W} {78, 084032}

\bibitem[\protect\citeauthoryear{{Wiltshire}}{{Wiltshire}}{2009}]{Wiltshire_2009_obs}
{Wiltshire} D.~L.,  2009, \mn@doi [\prd] {10.1103/PhysRevD.80.123512}, \href {https://ui.adsabs.harvard.edu/abs/2009PhRvD..80l3512W} {80, 123512}

\bibitem[\protect\citeauthoryear{Wiltshire}{Wiltshire}{2014}]{Wiltshire_2014_cosmic}
Wiltshire D.~L.,  2014, in {Perez Bergliaffa} S.,  {Novello} M.,  eds, , {Proceedings of the XVth Brazilian School of Cosmology and Gravitation}.
{Cambridge Scientific Publishers}, pp 203--244 (\mn@eprint {arXiv} {1311.3787})

\makeatother
\end{thebibliography}

\bsp	
\label{lastpage}
\end{document}